\documentclass[conference]{IEEEtran}
\usepackage{amsmath,amsfonts}
\usepackage{algorithmic}
\usepackage{algorithm}
\usepackage{array}
\usepackage[caption=false,font=normalsize,labelfont=sf,textfont=sf]{subfig}
\usepackage{textcomp}
\usepackage{stfloats}
\usepackage{url}
\usepackage{verbatim}
\usepackage{graphicx}
\usepackage{cite}
\usepackage{xcolor}

\begin{document}

\title{Electromagnetic-Compliant Channel Modeling and Performance Evaluation for Holographic MIMO}

\author{Tengjiao Wang\IEEEauthorrefmark{1}, Wei Han\IEEEauthorrefmark{1}, Zhimeng Zhong\IEEEauthorrefmark{1}, Jiyong Pang\IEEEauthorrefmark{1}, Guohua Zhou\IEEEauthorrefmark{1}, Shaobo Wang\IEEEauthorrefmark{1}, Qiang Li\IEEEauthorrefmark{2}
\\ \IEEEauthorrefmark{1}Wireless Network RAN Research Department, Huawei Technologies CO., Ltd, Shanghai, China 
\\ \IEEEauthorrefmark{2}Department of Broadband Communication, Peng Cheng Laboratory, Shenzhen, China
\\ Emails: \IEEEauthorrefmark{1}\{wangtengjiao6, wayne.hanwei, zhongzhimeng, pangjiyong, guohua.zhou, shaobo.wang\}@huawei.com, 
\\ \IEEEauthorrefmark{2}liq03@pcl.ac.cn
}        



\maketitle

\begin{abstract}
Recently, the concept of holographic multiple-input multiple-output (MIMO) is emerging as one of the promising technologies beyond massive MIMO. Many challenges need to be addressed to bring this novel idea into practice, including electromagnetic (EM)-compliant channel modeling and accurate performance evaluation. In this paper, an EM-compliant channel model is proposed for the holographic MIMO systems, which is able to model both the characteristics of the propagation channel and the non-ideal factors caused by mutual coupling at the transceivers, including the antenna pattern distortion and the decrease of antenna efficiency. Based on the proposed channel model, a more realistic performance evaluation is conducted to show the performance of the holographic MIMO system in both the single-user and the multi-user scenarios. Key challenges and future research directions are further provided based on the theoretical analyses and numerical results.
\end{abstract}


\section{Introduction}
In the past ten years, owing to the contributions from both academia and industry, massive multiple-input multiple-output (MIMO) technique has been well developed and greatly improved the performance of 5G cellular networks~\cite{2019-Bjornson-BeyondMM}. In the future 5G-Advanced and 6G cellular networks, new applications are rapidly emerging, which require much higher data rate, lower latency, and higher reliability~\cite{2022-Debbah-EIT}. 

In order to tackle these new challenges in the future wireless communication systems, the concept of holographic MIMO is proposed recently~\cite{2020-Chongwen-Holo}. 
By integrating an infinite number of antennas into a limited surface, holographic MIMO is expected to fully exploit the propagation characteristics offered by the electromagnetic channel and approach the fundamental performance limit~\cite{2021-Dardari-Holo}. Although it is an emerging technique,  lots of attention has been attracted from both academia and industry. In~\cite{2021-Linglong-CAP}, the capacity of a single-user holographic MIMO system is investigated. In~\cite{2022-Wei-DoF}, the degree of freedom of a single-user holographic MIMO system is analyzed. Then in~\cite{2022-Luca-HoloModel}, a channel estimation scheme is proposed to achieve higher accuracy for the holographic MIMO system.

In holographic MIMO, one of the key challenges is the accurate channel modeling. In~\cite{2020-Pizzo-HoloChannel}, a Fourier plane-wave series expansion-based channel model is proposed for holographic MIMO systems. Based on the Fourier spectral representation, it provides a physically meaningful model capturing the propagation characteristics of the electromagnetic (EM) wave. In~\cite{2022-Chongwen-HoloChannel}, the authors extend the Fourier plane-wave channel model to a multi-user scenario and  the achievable rate is further investigated. While the isotropic scattering environment is considered in~\cite{2020-Pizzo-HoloChannel, 2022-Chongwen-HoloChannel}, the non-isotropic scattering environment is further modeled in~\cite{2022-Pizzo-HoloChannel}. However, how to determine the non-isotropic angular power spectrum based on the realistic channel statistics remains an open problem~\cite{2022-Pizzo-HoloChannel}. Moreover, when the antenna elements are closely deployed, strong mutual coupling among the elements will cause non-ideal factors at the transceivers, including the distortion of antenna patterns and the decrease of antenna efficiency~\cite{2018-Xiaoming-MutualCouple, 2004-Jensen-MutualCouple}. These factors are not taken into consideration in the state-of-the-art channel models~\cite{2020-Pizzo-HoloChannel, 2022-Chongwen-HoloChannel, 2022-Pizzo-HoloChannel}, however, they will have significant impact on the performance of the holographic MIMO systems.




In this paper, we try to bridge this gap by proposing an EM-compliant channel model for holographic MIMO, which is able to model both the characteristics of the propagation channel and the non-ideal factors at the transceivers caused by antenna mutual coupling. The contributions of this paper can be summarized as follows:
\begin{itemize}
\item An EM-compliant channel model is proposed for the holographic MIMO systems. Based on the von Mises-Fisher (VMF) distributions~\cite{2009-Mammasis-Mises} and the 3GPP TR 38.901 channel model~\cite{2020-Standard-38901}, a realistic angular power spectrum is modeled, which has intuitive physical meaning and is more consistent with the realistic environment. 

\item The non-ideal factors caused by antenna mutual coupling at the transceivers are taken into consideration, including the antenna pattern distortion and the decrease of antenna efficiency. 

\item Using the proposed channel model, a more realistic performance evaluation is conducted in both the single-user and multi-user scenarios. Simulation results are provided to show the performance and technical challenges for the future holographic MIMO systems.

\end{itemize}

The rest of this paper is organized as follows. In Section~\ref{Sec-Conv}, the Fourier plane-wave series expansion-based channel model for holographic MIMO is briefly reviewed. In Section~\ref{Sec-Holo}, the proposed EM-compliant channel model is explained in detail. Then based on the proposed channel model, the performance of holographic MIMO is evaluated and the simulation results are provided in Section~\ref{Sec-Simu}. Finally, conclusions are drawn in Section~\ref{Sec-Conc}.

\textit{Notation:} Column vectors and matrices are denoted by lowercase and uppercase boldface letters.  Conjugate transposition is denoted by $(\cdot)^\mathrm{H}$. Cardinality of a set is denoted by $|\cdot|$. Frobenius norm is denoted by $\| \cdot \|$. The $p$-th element of a vector, and the element at the $p$-th row and the $q$-th column of a matrix are denoted by $[\cdot]_p$ and $[\cdot]_{p,q}$, respectively.


\begin{figure*}[!t]
\centering
\includegraphics[width=7in]{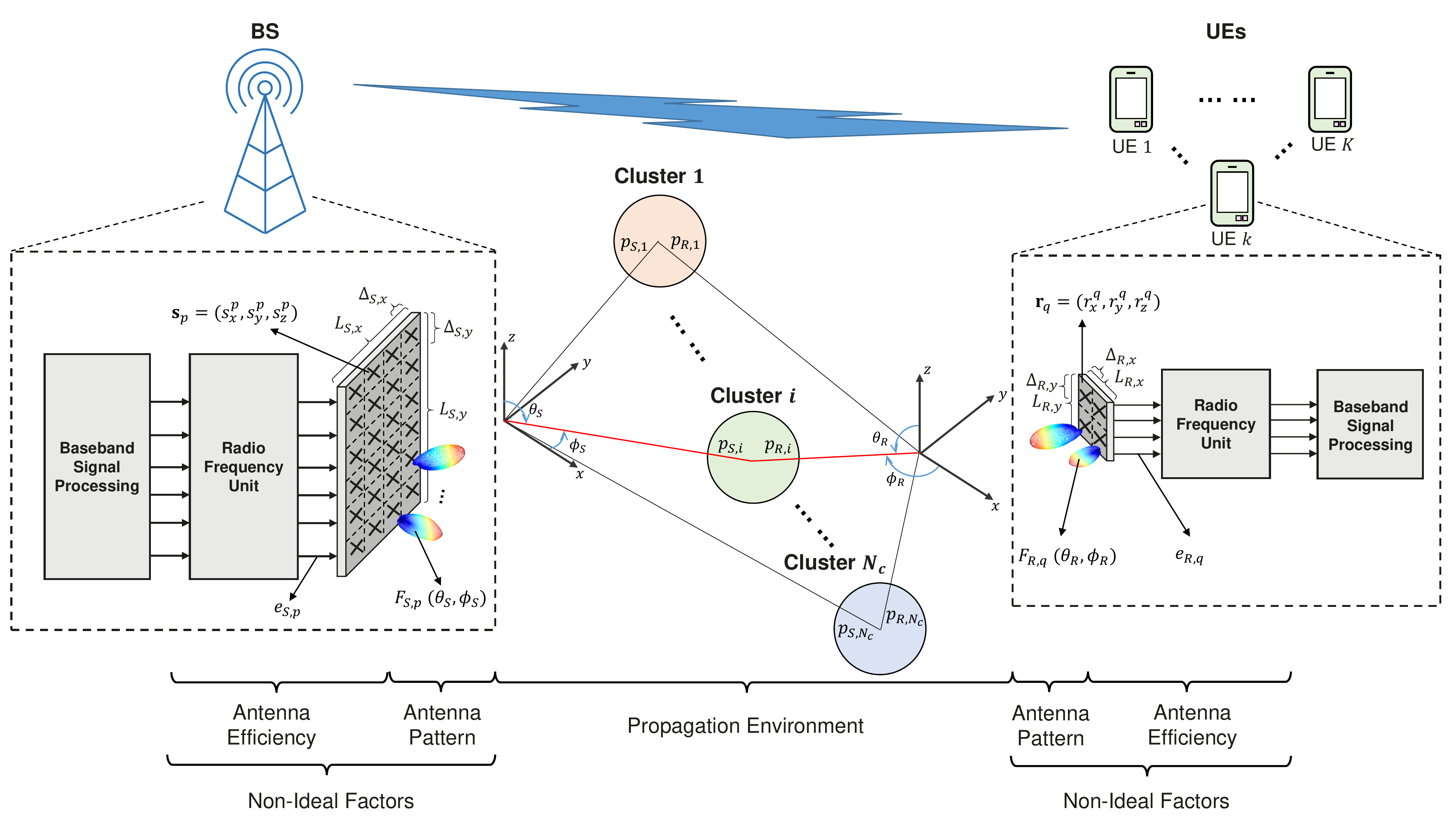}
\caption{Illustration of the proposed EM-compliant channel model for holographic MIMO systems, including the characteristics of the propagation environment and the non-ideal factors at the transceivers.}
\label{Fig-Struct}
\end{figure*}

\section{Fourier Plane-Wave Series Expansion-Based Channel Model} \label{Sec-Conv}
In this section, we briefly introduce the Fourier plane-wave series expansion-based channel model for holographic MIMO systems, which is a spatially-stationary small-scale fading channel model proposed in~\cite{2020-Pizzo-HoloChannel, 2022-Pizzo-HoloChannel,2022-Chongwen-HoloChannel}. 

We consider a holographic MIMO system equipped with uniform planar arrays at both the base station (BS) and the user equipment (UE). The width and height of the arrays at BS and UE are denoted by $\{ L_{\mathrm{S},x}, L_{\mathrm{S},y} \}$ and $\{ L_{\mathrm{R},x}, L_{\mathrm{R},y} \}$, respectively. $N_\mathrm{S}$ and $N_\mathrm{R}$ antenna elements are deployed with spacing $\{ \Delta_{\mathrm{S},x}, \Delta_{\mathrm{S},y} \}$ and $\{ \Delta_{\mathrm{R},x}, \Delta_{\mathrm{R},y} \}$ at BS and UE. The coordinates of the antenna elements are represented by $\mathbf{s}_p = (s^p_x, s^p_y, s^p_z), p = 1,2,\cdots,N_\mathrm{S}$ and $\mathbf{r}_q = (r^q_x, r^q_y, r^q_z), q = 1,2,\cdots,N_\mathrm{R}$, respectively. The central frequency and wavelength are denoted by $f_c$ and $\lambda$.

Based on the Fourier plane-wave series expansion, the channel matrix $\mathbf{H} \in \mathbb{C}^{N_\mathrm{R} \times N_\mathrm{S}}$ can be expressed as 
\begin{equation} \label{equ_H}
\begin{split}
	\mathbf{H} =& \sqrt{N_\mathrm{R} N_\mathrm{S}} \times \\
& \sum_{l_x,l_y}  \sum_{m_x,m_y} H_\mathrm{a}(l_x, l_y, m_x, m_y) \mathbf{a}_\mathrm{R}(l_x, l_y) \mathbf{a}_\mathrm{S}^\mathrm{H}(m_x, m_y),
\end{split}
\end{equation}
where $\mathbf{a}_\mathrm{S}(m_x, m_y)$ and $\mathbf{a}_\mathrm{R}(l_x, l_y)$ denote the phase-shifted 2-dimensional Fourier harmonics with
\begin{equation} \label{equ_a_S}
	[\mathbf{a}_\mathrm{S}(m_x, m_y)]_p = \!\! \frac{1}{\sqrt{N_\mathrm{S}}} e^{-j\left( \frac{2\pi m_x}{L_{\mathrm{S},x}} s^p_x + \frac{2\pi m_y}{L_{\mathrm{S},y}} s^p_y + \gamma_\mathrm{S}(m_x,m_y) s^p_z\right)},
\end{equation}
and 
\begin{equation} \label{equ_a_R}
	[\mathbf{a}_\mathrm{R}(l_x, l_y)]_q = \!\! \frac{1}{\sqrt{N_\mathrm{R}}} e^{j\left( \frac{2\pi l_x}{L_{\mathrm{R},x}} r^q_x + \frac{2\pi l_y}{L_{\mathrm{R},y}} r^q_y + \gamma_\mathrm{R}(l_x,l_y) r^q_z\right)},
\end{equation}
where $\gamma_\mathrm{S}(m_x, m_y) = \sqrt{(\frac{2\pi}{\lambda})^2 - (\frac{2\pi m_x}{L_{\mathrm{S},x}})^2 - (\frac{2\pi m_y}{L_{\mathrm{S},y}})^2}$ and $\gamma_\mathrm{R}(l_x, l_y) = \sqrt{(\frac{2\pi}{\lambda})^2 - (\frac{2\pi l_x}{L_{\mathrm{R},x}})^2 - (\frac{2\pi l_y}{L_{\mathrm{R},y}})^2}$.

In (\ref{equ_H}), $H_\mathrm{a}(l_x, l_y, m_x, m_y)$ is the random Fourier coefficient with $H_\mathrm{a}(l_x, l_y, m_x, m_y) \sim \mathcal{CN}(\sigma^2(l_x, l_y, m_x, m_y))$. The variance $\sigma^2(l_x, l_y, m_x, m_y)$ can be further derived as

\begin{equation}
\begin{split}
	\sigma^2(l_x, l_y, m_x, m_y) = & \int \!\!\! \int_{\Omega_\mathrm{R}(l_x, l_y)} \!\!\! A^2(\theta_\mathrm{R}, \phi_\mathrm{R}) \sin \theta_\mathrm{R} \mathrm{d}\theta_\mathrm{R} \mathrm{d}\phi_\mathrm{R} \times \\
& \int \!\!\! \int_{\Omega_\mathrm{S}(m_x, m_y)} \!\!\! A^2(\theta_\mathrm{S}, \phi_\mathrm{S}) \sin \theta_\mathrm{S} \mathrm{d}\theta_\mathrm{S} \mathrm{d}\phi_\mathrm{S},
\end{split}
\end{equation}
where $A^2(\theta_\mathrm{S}, \phi_\mathrm{S})$ and $A^2(\theta_\mathrm{R}, \phi_\mathrm{R})$ denote the angular power spectrum at BS and UE, respectively. $\{ \theta_\mathrm{S}, \phi_\mathrm{S} \}$ and $\{ \theta_\mathrm{R}, \phi_\mathrm{R} \}$ denote the elevation angle and the azimuth angle at BS and UE, respectively. $\Omega_\mathrm{S}(m_x, m_y)$ and $\Omega_\mathrm{R}(l_x, l_y)$ are the integration region, the details of which can be found in \cite{2020-Pizzo-HoloChannel}.

According to~\cite{2020-Pizzo-HoloChannel}, the Fourier coefficient $H_\mathrm{a}(l_x, l_y, m_x, m_y)$ is non-zero only within the lattice ellipse

\begin{equation}
	\mathcal{E}_\mathrm{S} = \left\{ (m_x, m_y) \in \mathbb{Z}^2: \left( \frac{m_x \lambda}{L_{\mathrm{S},x}} \right)^2 + \left( \frac{m_y \lambda}{L_{\mathrm{S},y}} \right)^2 \leq 1 \right\},
\end{equation}
and
\begin{equation}
	\mathcal{E}_\mathrm{R} = \left\{ (l_x, l_y) \in \mathbb{Z}^2: \left( \frac{l_x \lambda}{L_{\mathrm{R},x}} \right)^2 + \left( \frac{l_y \lambda}{L_{\mathrm{R},y}} \right)^2 \leq 1 \right\},
\end{equation}
at BS and UE, respectively. We denote the cardinalities of the sets as $n_\mathrm{S} = |\mathcal{E}_\mathrm{S} |$ and $n_\mathrm{R} = | \mathcal{E}_\mathrm{R} |$. Therefore, the channel matrix $\mathbf{H}$ in (\ref{equ_H}) can be further written as 
\begin{equation}
	\mathbf{H} = \mathbf{U}_\mathrm{R} {\mathbf{H}}_\mathrm{a} \mathbf{U}_\mathrm{S}^\mathrm{H},
\end{equation}
where ${\mathbf{H}}_\mathrm{a} \in \mathbb{C}^{n_\mathrm{R} \times n_\mathrm{S}}$ collects all the non-zero elements $H_\mathrm{a}(l_x, l_y, m_x, m_y)$. $\mathbf{U}_\mathrm{S} \in \mathbb{C}^{N_\mathrm{S} \times n_\mathrm{S}}$ collects the corresponding $n_\mathrm{S}$ Fourier harmonics defined in (\ref{equ_a_S}), and $\mathbf{U}_\mathrm{R} \in \mathbb{C}^{N_\mathrm{R} \times n_\mathrm{R}}$ collects the corresponding $n_\mathrm{R}$ Fourier harmonics defined in (\ref{equ_a_R}).

\section{EM-Compliant Channel Model} \label{Sec-Holo}
In the above section, the Fourier plane-wave series expansion-based channel model is briefly reviewed, which provides a physically-meaningful and tractable channel model for holographic MIMO systems~\cite{2020-Pizzo-HoloChannel, 2022-Pizzo-HoloChannel,2022-Chongwen-HoloChannel}. However, the angular power spectrum needs to be modeled according to realistic channel statistics. Moreover, the non-ideal factors at the transmitter and the receiver need to be considered. When the antenna elements are closely deployed, strong mutual coupling among the elements will arise. According to~\cite{2018-Xiaoming-MutualCouple}, the non-ideal factors caused by the mutual coupling include the distortion of antenna patterns and the decrease of antenna efficiency.

In this section, we try to bridge this gap by proposing an EM-compliant channel model based on the Fourier plane-wave series expansion-based model. In the proposed channel model, the angular power spectrum of the propagation environment, the distortion of antenna patterns, and the decrease of antenna efficiency are jointly considered to provide a more realistic channel model for the holographic MIMO systems. An illustration of the proposed model is shown in Fig.~\ref{Fig-Struct}.

\subsection{Angular Power Spectrum}
In the Fourier plane-wave series expansion-based channel model, the angular power spectrum $A^2(\theta_\mathrm{S}, \phi_\mathrm{S})$ and $A^2(\theta_\mathrm{R}, \phi_\mathrm{R})$ characterize the propagation of EM wave in the environment. In~\cite{2020-Pizzo-HoloChannel} and \cite{2022-Chongwen-HoloChannel}, only the isotropic propagation is considered with $A^2(\theta_\mathrm{S}, \phi_\mathrm{S}) = A^2(\theta_\mathrm{R}, \phi_\mathrm{R}) = 1$. In~\cite{2022-Pizzo-HoloChannel}, the non-isotropic environment is modeled by  the mixture of VMF distributions~\cite{2009-Mammasis-Mises}. However, only two clusters of scatters are considered, which is not consistent with the realistic channels. In this subsection, we combine the VMF distributions and the measurement results from the 3GPP TR 38.901 standard~\cite{2020-Standard-38901} to build a more realistic angular power spectrum for the channel model.

Suppose there are $N_c$ clusters of scatters in the environment between BS and UE. The angular power spectrum at BS and UE can be modeled by a mixture of $N_c$ VMF distributions
\begin{equation}
	A^2_\mathrm{S} (\theta_\mathrm{S}, \phi_\mathrm{S}) = \sum_{i=1}^{N_c} w_{\mathrm{S},i} p_{\mathrm{S},i}(\theta_\mathrm{S}, \phi_\mathrm{S}),
\end{equation}
and
\begin{equation}
	A^2_\mathrm{R} (\theta_\mathrm{R}, \phi_\mathrm{R}) = \sum_{i=1}^{N_c} w_{\mathrm{R},i} p_{\mathrm{R},i}(\theta_\mathrm{R}, \phi_\mathrm{R}),
\end{equation}
where $w_{\mathrm{S},i}$ and $w_{\mathrm{R},i}$ represent the normalization factor with $\sum_i^{N_c} w_{\mathrm{S},i} = \sum_i^{N_c} w_{\mathrm{R},i} = 1$. $p_{\mathrm{S},i}(\theta_\mathrm{S}, \phi_\mathrm{S})$ and $p_{\mathrm{R},i}(\theta_\mathrm{R}, \phi_\mathrm{R})$ represent the probability distribution function of the 3-dimensional VMF distribution, which can be further given by~\cite{2000-Mardia-Mises}
\begin{equation}
\begin{split}
	p_{\mathrm{S},i}&(\theta_\mathrm{S}, \phi_\mathrm{S}) = \frac{\alpha_{\mathrm{S},i}}{4 \pi \mathrm{sinh} (\alpha_{\mathrm{S},i})} \times \\
& e^{\alpha_{\mathrm{S},i}(\sin \theta_\mathrm{S} \sin \bar{\theta}_{\mathrm{S},i} \cos(\phi_\mathrm{S} - \bar{\phi}_{\mathrm{S},i}) + \cos \theta_\mathrm{S} \cos \bar{\theta}_{\mathrm{S},i})},
\end{split}
\end{equation}
and
\begin{equation}
\begin{split}
	p_{\mathrm{R},i}&(\theta_\mathrm{R}, \phi_\mathrm{R}) = \frac{\alpha_{\mathrm{R},i}}{4 \pi \mathrm{sinh} (\alpha_{\mathrm{R},i})} \times \\
& e^{\alpha_{\mathrm{R},i}(\sin \theta_\mathrm{R} \sin \bar{\theta}_{\mathrm{R},i} \cos(\phi_\mathrm{R} - \bar{\phi}_{\mathrm{R},i}) + \cos \theta_\mathrm{R} \cos \bar{\theta}_{\mathrm{R},i})},
\end{split}
\end{equation}
where $\{\bar{\phi}_{\mathrm{S},i}, \bar{\theta}_{\mathrm{S},i}\}$ and $\{\bar{\phi}_{\mathrm{R},i}, \bar{\theta}_{\mathrm{R},i}\}$ denote the elevation and azimuth angles of the $i$-th cluster at BS and UE, respectively. $\alpha_{\mathrm{S},i}$ and $\alpha_{\mathrm{R},i}$ denote the concentration parameters for the $i$-th cluster, which determine the angular spread (AS) of each cluster. AS is larger with smaller concentration parameter. When $\alpha_{\mathrm{S},i} = \alpha_{\mathrm{R},i} = 0$, it represent the isotropic propagation environment.

In 3GPP TR 38.901 standard~\cite{2020-Standard-38901}, the cluster delay line (CDL) model is defined for the link-level evaluations. In~\cite{2020-Standard-38901}, the azimuth angle of departure (AoD) $\phi_{i, \mathrm{AoD}}$, zenith angle of departure (ZoD) $\theta_{i, \mathrm{ZoD}}$, azimuth angle of arrival (AoA) $\phi_{i, \mathrm{AoA}}$, zenith angle of arrival (ZoA) $\theta_{i, \mathrm{ZoA}}$, angular spread of departure (ASD) $\delta_\mathrm{ASD}$, and angular spread of arrival (ASA) $\delta_\mathrm{ASA}$ are provided based on practical measurements. Therefore, it can be used as guidance to implement a more realistic angular power spectrum for the channel. It can be derived by setting $\bar{\phi}_{\mathrm{S},i} = \phi_{i, \mathrm{AoD}}$, $\bar{\theta}_{\mathrm{S},i} = \theta_{i, \mathrm{ZoD}}$, $\bar{\phi}_{\mathrm{R},i} = \phi_{i, \mathrm{AoA}}$, and $\bar{\theta}_{\mathrm{R},i} = \theta_{i, \mathrm{ZoA}}$. The relation between AS and the concentration parameters in VMF can be found in Appendix 3.1 in~\cite{2000-Mardia-Mises}. For small AS $\delta_\mathrm{ASD} < 21^{\circ}$ and $\delta_\mathrm{ASA} < 21^{\circ}$, the concentration parameter can be derived by~\cite{2000-Mardia-Mises}
\begin{equation}
	\alpha_{\mathrm{S},i} = \frac{212.9^2}{\delta_\mathrm{ASD}^2},~\alpha_{\mathrm{R},i} = \frac{212.9^2}{\delta_\mathrm{ASA}^2}
\end{equation}

\begin{figure}[!t]
\centering
\includegraphics[width=3.5in]{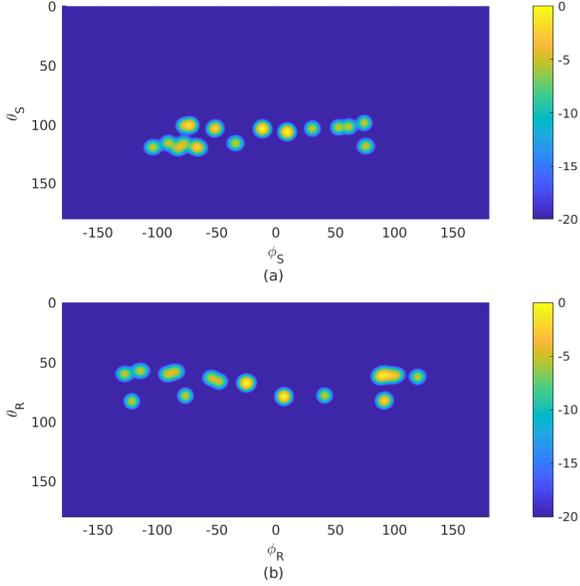}
\caption{Illustration of the anglular power spectrum of the propagation channel based on 3GPP CDL-B channel model. a)~Anglular power spectrum at BS; b)~Anglular power spectrum at UE.}
\label{Fig-Angle}
\end{figure}

For example, when refering to the CDL-B channel in~\cite{2020-Standard-38901}, the number of clusters should be set to $N_c = 23$. The angles AoD, ZoD, AoA, ZoA, ASD, and ASA of each cluster can be found in Table 7.7.1-2 in~\cite{2020-Standard-38901}. Based on the proposed model, the overall angular power spectrum can be illustrated in Fig.~\ref{Fig-Angle}.

\subsection{Antenna Pattern Distortion}

When the antenna elements are closely spaced, the mutual coupling among the antenna elements will distort the antenna pattern of each individual element~\cite{2018-Xiaoming-MutualCouple}. Firstly, the distorted antenna pattern will be different from that of an isolated element. Secondly, the distorted antenna patterns of different elements will also be different based on the locations of the elements in the array~\cite{2018-Xiaoming-MutualCouple}. 

In the previous works of channel modeling for holographic MIMO~\cite{2020-Pizzo-HoloChannel, 2022-Pizzo-HoloChannel,2022-Chongwen-HoloChannel}, this effect is neglected by assuming each element has a uniform antenna pattern. In this subsection, we involve the distortion of antenna pattern into the channel modeling of holographic MIMO systems.

We denote the normalized embedded antenna pattern of each individual antenna at BS and UE as $F_{\mathrm{S},p}(\theta_\mathrm{S}, \phi_\mathrm{S})$ and $F_{\mathrm{R},q}(\theta_\mathrm{R}, \phi_\mathrm{R})$. For specific antenna arrays at BS and UE, the antenna pattern can be obtained from electromagnetic computing softwares. In the Fourier plane-wave series expansion-based channel model, each Fourier harmonic $\mathbf{a}_\mathrm{S}(m_x, m_y)$ and $\mathbf{a}_\mathrm{R}(l_x, l_y)$ denotes a plane wave from a specific angle. Based on this, a modified Fourier harmonic is proposed to take the antenna pattern of individual element into consideration, which can be expressed as
\begin{equation} \label{equ_phi_S}
\begin{split}
	[\boldsymbol{\psi}_\mathrm{S}(m_x, m_y)]_p =
	 &\frac{1}{\sqrt{N_\mathrm{S}}} e^{-j\left( \frac{2\pi m_x}{L_{\mathrm{S},x}} s^p_x + \frac{2\pi m_y}{L_{\mathrm{S},y}} s^p_y + \gamma_\mathrm{S}(m_x,m_y) s^p_z\right)} \\
 	& \times F_{\mathrm{S},p}\left( \theta^*_\mathrm{S}(m_x, m_y) , \phi^*_\mathrm{S}(m_x, m_y) \right),
\end{split}
\end{equation}
and
\begin{equation} \label{equ_phi_R}
\begin{split}
	[\boldsymbol{\psi}_\mathrm{R}(l_x, l_y)]_q =
	& \frac{1}{\sqrt{N_\mathrm{R}}} e^{j\left( \frac{2\pi l_x}{L_{\mathrm{R},x}} r^q_x + \frac{2\pi l_y}{L_{\mathrm{R},y}} r^q_y + \gamma_\mathrm{R}(l_x,l_y) r^q_z\right)} \\
	& \times F_{\mathrm{R},q}\left( \theta^*_\mathrm{R}(l_x, l_y) , \phi^*_\mathrm{R}(l_x, l_y) \right),
\end{split}
\end{equation}
where~$\{ \theta^*_\mathrm{S}(m_x, m_y) , \phi^*_\mathrm{S}(m_x, m_y) \}$~and~$\{ \theta^*_\mathrm{R}(l_x, l_y) , \phi^*_\mathrm{R}(l_x, l_y) \}$ denote the corresponding elevation and azimuth angles for the Fourier harmonics $(m_x,m_y)$ and $(l_x,l_y)$ at BS and UE. These angles can be further given by
\begin{equation}
	\theta^*_\mathrm{S}(m_x, m_y) = \mathrm{arccos} \left( \sqrt{1 - \left(\frac{m_x \lambda}{L_{\mathrm{S},x}}\right)^2 - \left(\frac{ m_y \lambda}{L_{\mathrm{S},y}}\right)^2}  \right), 
\end{equation}
\begin{equation}
	\phi^*_\mathrm{S}(m_x, m_y) = \mathrm{arctan}\left( \frac{L_{\mathrm{S},x}m_y }{L_{\mathrm{S},y}m_x} \right),
\end{equation}
\begin{equation}
	\theta^*_\mathrm{R}(l_x, l_y) = \mathrm{arccos} \left( \sqrt{1 - \left(\frac{l_x \lambda}{L_{\mathrm{R},x}}\right)^2 - \left(\frac{ l_y \lambda}{L_{\mathrm{R},y}}\right)^2}  \right), 
\end{equation}
\begin{equation}
	\phi^*_\mathrm{R}(l_x, l_y) = \mathrm{arctan}\left( \frac{L_{\mathrm{R},x}l_y }{L_{\mathrm{R},y}l_x} \right).
\end{equation}
When the uniform antenna pattern is used, which implies $F_{\mathrm{S},p}(\theta_\mathrm{S}, \phi_\mathrm{S}) = F_{\mathrm{R},q}(\theta_\mathrm{R}, \phi_\mathrm{R}) = 1$, the modified Fourier harmonics $\boldsymbol{\psi}_\mathrm{S}(m_x, m_y)$ and $\boldsymbol{\psi}_\mathrm{R}(l_x, l_y)$ will be reduced to the conventional Fourier harmonics $\mathbf{a}_\mathrm{S}(m_x, m_y)$ and $\mathbf{a}_\mathrm{R}(l_x, l_y)$ in (\ref{equ_a_S}) and (\ref{equ_a_R}), respectively.

Therefore, the overall channel matrix can be written as 
\begin{equation}
	\mathbf{H} = \boldsymbol{\Psi}_\mathrm{R} {\mathbf{H}}_\mathrm{a} \boldsymbol{\Psi}_\mathrm{S}^\mathrm{H},
\end{equation}
where $\boldsymbol{\Psi}_\mathrm{S} \in \mathbb{C}^{N_\mathrm{S} \times n_\mathrm{S}}$ and $\boldsymbol{\Psi}_\mathrm{R} \in \mathbb{C}^{N_\mathrm{R} \times n_\mathrm{R}}$ collect the modified Fourier harmonics with antenna pattern distortion defined in (\ref{equ_phi_S}) and (\ref{equ_phi_R}), respectively.

\subsection{Antenna Efficiency}

In a dense antenna array with small element spacing, the efficiency of the antenna elements will be decreased because of the mutual coupling among them~\cite{2018-Xiaoming-MutualCouple}. The reason behind this is that due to the effect of mutual coupling, the active impedance of each element varies with scan angles. The impedance mismatch between the antenna elements and the transmission lines causes the power to be returned to the generators. Therefore, the efficiency is reduced~\cite{1964-Hannan-Limit}.

We denote the S-parameter matrix of the antenna arrays at BS and UE as $\mathbf{S}_\mathrm{S} \in \mathbb{C}^{N_\mathrm{S} \times N_\mathrm{S}}$ and $\mathbf{S}_\mathrm{R} \in \mathbb{C}^{N_\mathrm{R} \times N_\mathrm{R}}$. For a specific antenna array, this can be obtained from the electromagnetic computing softwares. According to~\cite{2015-Kidal-Limit}, the efficiency of each antenna element can be estimated as
\begin{equation}
	e_{\mathrm{S},p} =  1 - \sum_{p' = 1}^{N_\mathrm{S}} \|[\mathbf{S}_\mathrm{S}]_{p,p'}\|^2,
\end{equation}
and
\begin{equation}
	e_{\mathrm{R},q} =  1 - \sum_{q' = 1}^{N_\mathrm{R}} \|[\mathbf{S}_\mathrm{R}]_{q,q'}\|^2,
\end{equation} 
where $e_{\mathrm{S},p}$ and $e_{\mathrm{R},q}$ represent the efficiency of the $p$-th element at BS and $q$-th element at UE.

In~\cite{2015-Kidal-Limit, 1964-Hannan-Limit}, a relationship between the element efficiency and the element spacing is established, which is called Hannan's efficiency. It shows that the maximum available efficiency of an element in a dense array can be expressed as
\begin{equation}
	e_{\mathrm{S},p}^*(\Delta_{\mathrm{S},x}, \Delta_{\mathrm{S},y}) =  \frac{\pi L_{\mathrm{S},x} L_{\mathrm{S},y}}{N_\mathrm{S}\lambda^2} = \frac{\pi \Delta_{\mathrm{S},x} \Delta_{\mathrm{S},y}}{\lambda^2},
\end{equation}
and
\begin{equation}
	e_{\mathrm{R},q}^*(\Delta_{\mathrm{R},x}, \Delta_{\mathrm{R},y}) =  \frac{\pi L_{\mathrm{R},x} L_{\mathrm{R},y}}{N_\mathrm{R}\lambda^2} = \frac{\pi \Delta_{\mathrm{R},x} \Delta_{\mathrm{R},y}}{\lambda^2},
\end{equation}
which mean that the element efficiency is proportional to the area allocated to the element. In order to compare the performance of the holographic MIMO with different antenna spacing, we define a relative efficiency parameter $\eta_{\mathrm{S},p}$ and $\eta_{\mathrm{R},q}$ to account for the relative element efficiency compared to the classic array with half-wavelength element spacing. They can be expressed as
\begin{equation}
	\eta_{\mathrm{S},p} = \frac{e_{\mathrm{S},p}}{e^*_{\mathrm{S},p}(\frac{\lambda}{2}, \frac{\lambda}{2})},
\end{equation}
and
\begin{equation}
	\eta_{\mathrm{R},q} = \frac{e_{\mathrm{R},q}}{e^*_{\mathrm{R},q}(\frac{\lambda}{2}, \frac{\lambda}{2})}.
\end{equation}

Finally, when the antenna efficiency of each antenna element at BS and UE is accounted for, the overall channel matrix can be given by
\begin{equation} \label{H_final}
	\mathbf{H} = \boldsymbol{\Gamma}_\mathrm{R} \boldsymbol{\Psi}_\mathrm{R} {\mathbf{H}}_\mathrm{a} \boldsymbol{\Psi}_\mathrm{S}^\mathrm{H} \boldsymbol{\Gamma}_\mathrm{S},
\end{equation}
where $\boldsymbol{\Gamma}_\mathrm{S} \in \mathbb{R}^{N_\mathrm{S} \times N_\mathrm{S}}$ and $\boldsymbol{\Gamma}_\mathrm{R} \in \mathbb{R}^{N_\mathrm{R} \times N_\mathrm{R}}$ are diagonal matrices indicate the efficiency of each antenna element with $[\boldsymbol{\Gamma}_\mathrm{S}]_{p,p} = e_{\mathrm{S},p} = \eta_{\mathrm{S},p} e^*_{\mathrm{S},p}(\frac{\lambda}{2}, \frac{\lambda}{2})$ and 
$[\boldsymbol{\Gamma}_\mathrm{R}]_{q,q} = e_{\mathrm{R},q} = \eta_{\mathrm{R},q} e^*_{\mathrm{R},q}(\frac{\lambda}{2}, \frac{\lambda}{2})$.

\section{Numerical Results}	\label{Sec-Simu}


In this section, the performance of the holographic MIMO system is evaluated based on the proposed EM-compliant channel model. The downlink channel capacities of the single-user and multi-user holographic MIMO systems are investigated in the following subsections.

\subsection{Single-User Scenario}

\begin{figure}[!t]
\centering
\includegraphics[width=3.5in]{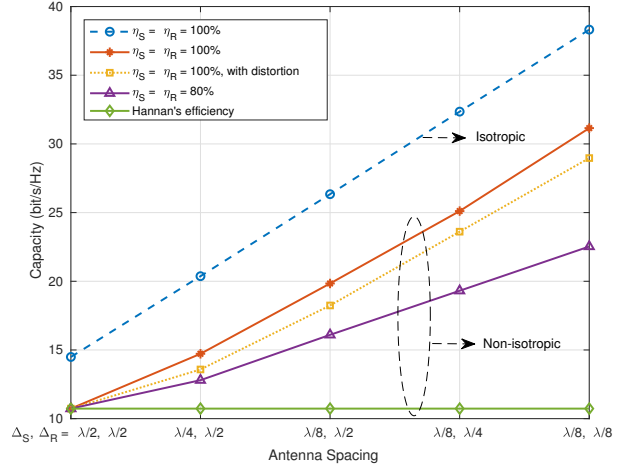}
\caption{Ergodic capacity of a single-user holographic MIMO system with different antenna spacing at BS and UE. }
\label{Res-1}
\end{figure}

Firstly, the ergodic capacity with different antenna spacing at BS and UE is analyzed. The central frequency is set to be $f_c = 3.5$~GHz. The width and height of the antenna arrays at BS and UE are fixed as $\{ L_{\mathrm{S},x}, L_{\mathrm{S},y}\} = \{4 \lambda, 4 \lambda\}$ and $\{ L_{\mathrm{R},x}, L_{\mathrm{R},y}\} = \{\lambda, \lambda\}$, respectively. Equal element spacing is adopted in the width and height with $\Delta_\mathrm{S} = \Delta_{\mathrm{S},x} = \Delta_{\mathrm{S},y}$ and $\Delta_\mathrm{R} = \Delta_{\mathrm{R},x} = \Delta_{\mathrm{R},y}$. The number of antenna elements can be calculated by $N_\mathrm{S} = L_{\mathrm{S},x}L_{\mathrm{S},y}/\Delta_{\mathrm{S}}^2$ and $N_\mathrm{R} = L_{\mathrm{R},x}L_{\mathrm{R},y}/\Delta_{\mathrm{R}}^2$. The proposed EM-compliant channel model in (\ref{H_final}) is utilized. In the non-isotropic scattering environment, the angular power spectrum of VMF distribution is generated based on the CDL-B channel in 3GPP TR 38.901~\cite{2020-Standard-38901} with $N_c = 23$ clusters. In the isotropic scattering environment, the parameters are set to be $N_c = 1$ and $\alpha_{\mathrm{S},i} = \alpha_{\mathrm{R},i} = 0$. When the antenna pattern distortion is considered, it is generated from the electromagnetic computing softwares where each element is modeled as a dipole antenna. Water filling power allocation strategy is applied to maximize the single-user capacity. The signal to noise ratio (SNR) is set to be 0~dB. 

The ergodic capacity is analyzed through Monte Carlo simulations with 1000 channel realizations. Three main findings can be summarized from the numerical results shown in Fig.~\ref{Res-1}.
\begin{itemize}
\item The characteristics of the scattering environment affect the channel capacity. It is shown that the capacity in the isotropic environment (dashed line) is much larger than that in the non-isotropic environment (solid line) with the same antenna spacing. The main reason is that the angular spread in the isotropic environment is much larger, which can provide more spatial degrees of freedom than the non-isotropic environment.
 
\item The antenna efficiency has significant impact on the channel capacity of holographic MIMO. When the closely spaced antenna elements have $100\%$ relative efficiency, which means $\eta_{\mathrm{S},q} = \eta_{\mathrm{R},q} = 100\%$, integrating more antenna elements into the limited antenna aperture can increase the channel capacity. For example, when BS and UE are equipped with $\lambda/8$ spacing elements, it can achieve about $300\%$ capacity gain. However, when the efficiency deteriorates to $80\%$, the capacity gain will reduce to about $200\%$. It is worth noting that when the efficiency is limited by the area allocated to the element according to Hannan's efficiency~\cite{1964-Hannan-Limit}, integrating more antenna elements cannot provide any capacity gain, which is because the array gain and multiplexing gain by deploying more antenna elements will be completely eaten by the decrease of the antenna efficiency.

\item The antenna pattern distortion may also decrease the channel capacity. It can be seen that when the distorted antenna pattern is applied (dotted line), the channel capacity will decrease about $5\%$ to $10\%$ at the same antenna efficiency.
\end{itemize}

\subsection{Multi-User Scenario}

Then, we further investigate the ergodic capacity of the multi-user scenario. We consider a cellular sector with one BS and $K=10$ users. The users are uniformly distributed in the sector within the angle of $[-120^{\circ}, 120^{\circ}]$ and the distance of $[25,100]$ meters. The large-scale pathloss is modeled according to the urban macro (UMa) scenario in 3GPP TR 38.901~\cite{2020-Standard-38901}. The SNR for the user which is 50 meters from BS is normalized to 0~dB. The iterative water-filling strategy is adopted to calculate the multi-user sum capacity~\cite{2006-Wei-WaterFill}.

\begin{figure}[!t]
\centering
\includegraphics[width=3.5in]{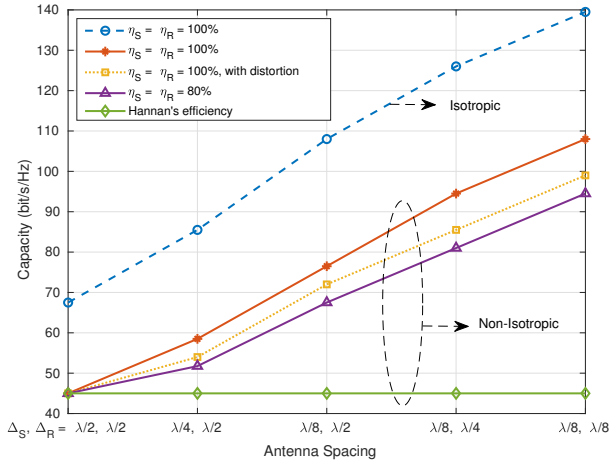}
\caption{Ergodic capacity of a multi-user holographic MIMO system with different antenna spacing at BS and UEs. }
\label{Res-2}
\end{figure}

Similar conclusions can be drawn from the simulation results in Fig.~\ref{Res-2}. When only the impact of the EM propagation environment is considered, holographic MIMO is able to increase the ergodic capacity when more closely spaced antenna elements are used. However, the non-ideal factors at the transceivers will prevent us from obtaining the whole capacity gain. These non-ideal factors have posed the following challenges for designing a holographic MIMO system:
\begin{itemize}
\item Antenna Efficiency: The efficiency of the antenna element is decreased because of mutual coupling and impedance mismatch. Adaptive impedance matching networks and super-gain antennas may be possible ways to increase the element efficiency.

\item Antenna Pattern Distortion: The distortion of individual antenna patterns has negative impact of the beamforming effect, which affect the overall capacity of the system.

\item Channel Correlation: The angular power spectrum determines the channel correlation. Smart environment control techniques may be needed to decrease the channel correlation and increase the capacity.
\end{itemize}
These challenges should be carefully dealt with in the design of holographic MIMO systems.

\vspace{-0.2cm}


\section{Conclusion} \label{Sec-Conc}
In this paper, an EM-compliant channel model has been proposed for the future holographic MIMO communication systems, which takes the characteristics of the scattering environment, antenna pattern distortion, and antenna efficiency into consideration. Based on the proposed channel model, numerical simulations have been conducted to show the performance of the holographic MIMO systems. Key challenges are also highlighted to inspire further research into this fast growing area.

\bibliographystyle{IEEEtran}
\bibliography{MyBib}



\vfill

\end{document}